\documentclass[pra,twocolumn,showpacs,nofootinbib,10pt]{revtex4-1} 
\usepackage{amsmath,amssymb,graphicx,color}

\usepackage[draft,implicit=false]{hyperref}
\usepackage{amsmath,amssymb,graphicx,color}

\newcommand{\bc}{\begin{cases}\begin{aligned}}
\newcommand{\ec}{\end{aligned}\end{cases}}
\newcommand{\eq}{\begin{equation}}
\newcommand{\fine}{\end{equation}}
\newcommand{\beq}{\begin{equation}}
\newcommand{\eeq}{\end{equation}}

\newcommand{\uno}{\leavevmode\hbox{\small1\normalsize\kern-.33em1}}
\newcommand{\xx}{{\bf x}}

\newcommand{\dd}{{\rm d}}

\newcommand{\casi}{\begin{cases}\begin{aligned}}
\newcommand{\casiend}{\end{aligned}\end{cases}}

\newcommand{\re}{\Re {\rm  e}}
\newcommand{\im}{\Im {\rm  m}}

\newcommand{\CiB}{{\rm CiB}_{p,m}^{(q_0,q_1)}(\xx)}
\newcommand{\Ftwoone}{\Psi_{p,m}^{(\xi)}}

\newcommand{\thrms}{\theta_{\rm rms}}
\newcommand{\thEE}{\theta_{\rm EE}}

\begin{document}


\title{On the properties of Circular-Beams: normalization, Laguerre-Gauss expansion and free-space divergence}

\author{Giuseppe Vallone}
\affiliation{Department of Information Engineering, University of Padova, via Gradenigo 6/B, 35131 Padova, Italy}

\begin{abstract}
Circular-Beams were introduced as a very general solution of the paraxial wave equation carrying Orbital Angular Momentum.
Here we study their properties, by looking at their normalization and their expansion in terms of Laguerre-Gauss modes.
We also study their far-field divergence and, for particular cases of the beam parameters, their possible experimental generation.
\end{abstract}


\maketitle 


Orbital Angular Momentum (OAM)~\cite{alle92pra}  of light has recently attracted a lot of attention 
as a new promising resource for fundamental and applied physics~\cite{grie03nat,
furh05ope,tamb11nph, damb12nco, bozi13sci, damb13nco,
damb13prx, lave13sci, grie03nat,vall14prl}, 
such as biophysics~\cite{grie03nat}, microscopy~\cite{furh05ope}, astronomy
~\cite{tamb11nph} and  metrology~\cite{damb13nco}. 
In this framework, Circular-Beams (CiBs) were introduced in~\cite{band08opl}
as a very general  solution carrying OAM of the paraxial wave equation.
The importance of the CiBs is underlined by the fact that they represent
a generalization of many well known  beams with OAM. Indeed, for particular values of the beam parameters, the CiBs reduce to 
standard~\cite{sieg86lasers} or elegant Laguerre-Gauss~\cite{wuns89josa}, Hypergeometric-Gauss~\cite{kari07opl,kari09opl},
 Hypergeometric~\cite{kotl07opl} modes and others~\cite{band08opl}. Then, the study of the CiB properties allows
 to better understand, in a unified way, the features of such well known beams.
For instance, since the propagation of CiBs through ABCD optical systems can be easily described in terms of their defining parameters
$q_0$ and $q_1$~\cite{band08opl}, the same  propagation law can be also applied for their known particular cases. 
 Moreover, while Laguerre-Gauss (LG) beams seem a natural choice for describing OAM states in relatively simple terms, in several cases
they only represent an approximation of the generated beam.
Indeed, a special CiB quite different from an LG mode, 
 is obtained by applying a singular phase factor $\exp(i m\phi)$ to a Gaussian beam (see eq. \eqref{pupil_limit} and~\cite{kari07opl}).
This technique represents a very common and simple way to generate OAM light~\cite{damb12nco,damb13prx,vall14prl}.
Thus, the study of the Circular-Beams  gives insight for the precise modeling, 
generation and exploitation of OAM in classical or quantum applications.

In this letter, we analyze the properties of the CiBs: in particular, we will explicitly derive 
their expansion in terms of  Laguerre-Gauss modes and their normalization constant, 
allowing the evaluation of their free-space divergence.
We will also investigate the constraints on the values of the beam parameters arising by the request of 
square integrability and regularity at any finite point. Then, we will 
study the experimental generation of a subclass of CiBs. Our results will provide an useful tool for the complete modeling 
of CiBs, paving the way towards their experimental implementation.   

CiBs were introduced in~\cite{band08opl}, without the evaluation of their normalization. 
As a first result of this letter, we give the detailed expression of the normalized CiB.
 Let's define the complex parameters $q(z), \widetilde q(z)$, $\xi$ and the scale factor $\chi(z)$ as:
\begin{align}
\notag q(z)&=z+q_0\,,\quad  &\xi&=\frac{q_1-q_0}{q^*_0-q_1}\,,
\\
\widetilde q(z)&=z+q_1\,, &\frac{1}{\chi^2(z)}&=\frac{ik}{2}\left[\frac1{q(z)}-\frac1{\widetilde q(z)}\right]\,,
\end{align}
were the real and imaginary part of the two complex beam parameter $q_0$ and $q_1$ are given by 
$q_0=iz_0-d_0$, $q_1=iz_1-d_1$ being $z_0=\pi W^2_0/\lambda$ and $d_0$
the analogues of the confocal parameter and the location of the waist for a Gaussian beam. 
The normalized monochromatic CiB in cylindrical coordinates $\xx\equiv(r,\phi,z)$ and propagating along the $z$ axis 
is defined as 
$\exp[i(\omega t-k z)]{\rm CiB}_{p,m}^{(q_0,q_1)}(\xx)$ where
%
\begin{align}
\label{cib}
\notag\CiB=&{(i\sqrt{2}\frac{z_0}{W_0})}^{|m|+1}\left[\pi\, |m|!\,\Ftwoone\right]^{-\frac12}\times
\\\
\notag&\frac{e^{-\frac{ikr^2}{2q(z)}}}{q(z)}
{\left[(1+\xi)\frac{\widetilde q(z)}{q(z)}\right]}^{\frac {p}2}
{\left[\frac{r}{q(z)}\right]}^{{|m|}}
\times
\\
&
\ _1F_1(-\frac p2,|m|+1;\frac{r^2}{\mathcal \chi^2(z)})e^{im\phi}\,.
\end{align}
\noindent 
In the above equation $_1F_1$ is the confluent Hypergeometric function 
$ _1F_1(a,b;x)=\sum^{+\infty}_{n=0}\frac{\Gamma(a+n)\Gamma(b)}{\Gamma(a)\Gamma(b+n)}\frac{x^n}{n!}$ and
$\Ftwoone$ a normalization factor given by
\begin{align}
\notag
\Ftwoone&=\sum^{+\infty}_{n=0}\frac{\Gamma(n-\frac p2)\Gamma(n-\frac {p^*}2)|m|!}
{\Gamma (-\frac p2)\Gamma (-\frac {p^*}2)n!(|m|+n)!}
|\xi|^{2n}
\\&\equiv\!\ _2F_1[-\frac p2, -\frac{p^*}{2}, 1 + |m|,|\xi|^2]\,.
\label{F21}
\end{align}
The parameters $m\in \mathbb Z$ and $p\in\mathbb C$ respectively represent 
 the amount of OAM carried by the beam and the analogous of the Laguerre-Gauss  mode radial index.
As noted in~\cite{band08opl}, the field is invariant (up to a normalization factor) by the transformation
$(p,m, q_0,q_1)\rightarrow (-p-2|m|-2,m, q_1, q_0)$\footnote{In~\cite{band08opl} the field was defined with $p$ replaced by $i\gamma-|m|-1$}.
As we will see, to guarantee the square integrability and regularity at each finite $z$ and $r$, 
some constraints will arise on the allowed values of $p$.

As a second result of this letter, we give 
 the expansion of the CiB in terms of the 
 Laguerre-Gauss beams ${\rm LG}_{n,m}(\xx)$ when $q_1\neq q^*_0$ and $\im(q_0)>0$:
\begin{align}
{\rm CiB}_{p,m}^{(q_0,q_1)}(\xx)
&=
\sum^{+\infty}_{n=0}{A}^{(m,\xi)}_{p,n}
{\rm LG}_{n,m}(r,\phi,z-d_0)\,,
\label{LGexpansion}
\end{align}
where
\begin{align}
{A}^{(m,\xi)}_{p,n}&=\frac{\xi^n}{
\sqrt{\Ftwoone}}
\frac{\Gamma(n-\frac p2)}{\Gamma (-\frac p2)}
\sqrt{\frac{|m|!}{n!(|m|+n)!}}\,.
\end{align}
The above expansion does not hold when $q_1= q^*_0$ and $\im(q_0)>0$: indeed, in this case, 
the LG$_{n,m}$ mode is directly obtained as
 a particular CiB, namely
${\rm LG}_{n,m}(r,\phi,z-d_0)=(-1)^n{\rm CiB}_{2n,m}^{(q_0,q^*_0)}(\xx)$.
Equation \eqref{LGexpansion} implicitly shows that
the CiBs satisfy the paraxial wave equation
since they can be represented as 
a linear combination of its solutions (i.e. the LG modes). 
For completeness, we here report the explicit expression of the standard Laguerre-Gauss modes:
\begin{align}
\notag{\rm LG}_{n,m}(\xx)=&\sqrt{\frac{2}{\pi}}
\sqrt{\frac{ n!}{(|m|+n)!}}
\frac{e^{-\frac{ikr^2}{2q(z)}}}{W(z)}{\left[\frac{\sqrt{2}r}{W(z)}\right]}^{|m|}
\times\\
&L_n^{(|m|)}\!\!\left[\frac{2 r^2}{W^2(z)}\right]e^{im\phi}e^{i(2n+|m|+1)\zeta(z)}\,,
\end{align}
where $n$, $m\in \mathbb Z$ with $n\geq 0$, $L_n^{(|m|)}(x)$ is the generalized Laguerre polynomial,
$W(z)=W_0\sqrt{1+(z/z_0)^2}$ the beam size and
$\exp[i\zeta(z)]=(z_0+i z)/|z_0+i z|$ the Gouy phase. 

We now demonstrate eqs. \eqref{F21} and \eqref{LGexpansion}.
When $q_1\neq q^*_0$ and $\im(q_0)>0$
we may exploit the expansion of the Hypergeometric function in terms of Laguerre polynomials:
\beq
\! _1F_1(a,b,\frac{Y Z}{Y-1}) = \!\sum^{+\infty}_{n=0}\frac{\Gamma(a+n)\Gamma (b)}{\Gamma (a)\Gamma (b+n)}\frac{Y^nL_n^{(b-1)}(Z)}{(1-Y)^{-a}}.
\eeq
Indeed, 
by using $Z=\frac{2r^2}{W(z-d_0)^2}$ and $Y=-\xi\frac{q(z)^*}{q(z)}$ in the above equation,  relation
 \eqref{LGexpansion} can be easily obtained.
Regarding the value of the constant  $\Ftwoone$ (see eq. \eqref{F21}), it can be determined directly from the expansion given in
 eq.  \eqref{LGexpansion}.
By recalling that the Laguerre-Gauss modes are orthonormal with respect to the scalar product
$
\int {\rm d}\phi\int{\rm d}r\,r\, {\rm LG}^*_{n_1,m_1}({\bf x}){\rm LG}_{n_2,m_2}({\bf x})=~\delta_{n_1,n_2}\delta_{m_1,m_2}$,
the CiB is easily found to be normalized to 1
if the constant $\Ftwoone$ is defined as \eqref{F21}.

As already anticipated, to obtain square integrable and regular beams, 
some constraints (summarized in eq. \eqref{conditions}) arises on the allowed values of $p$.
Let's start by the square integrability,
equivalent to the convergence of the sum in eq. \eqref{F21}. The latter converges to the Hypergeometric function
$_2F_1$ in three cases: 
a) when $\im(q_0)>0$, $\im(q_1)>0$, $\forall p$; 
b) when $\im(q_0)>0$, $\im(q_1)=0,+\infty$ and $\re(p)>-1-|m|$; c) when $\im(q_0)>0$, $\im(q_1)<0$
and $p=2\ell$ with $\ell\in \mathbb N$. In the latter case the sum becomes finite, 
since in eq. \eqref{LGexpansion} the expression $\sum^{+\infty}_{n=0}{\Gamma(n-\frac p2)}/{\Gamma (-\frac p2)}$
must be replaced by  $\sum^{\ell}_{n=0}{(-1)^n\ell!}/{(\ell-n)!}$. 
The three  cases above cited respectively correspond  to $|\xi|<1$,
$|\xi|=1$ and $|\xi|>1$.
As noticed in ~\cite{band08opl}, the same constraints can be obtained by looking at the CiB behavior at large $r$.
%

However, further conditions on $p$ arise by requiring the regularity of the field at each finite $z$. 
 In the cases a) and c) the field is regular at any finite $r$ and $z$.
On the other hand, when $\im(q_0)>0$ and $\im(q_1)=0$ the field may be singular at $z=d_1$.
Indeed, since in this case we have
$\lim_{z\rightarrow d_1}
\CiB
\propto 
r^{p+|m|}$, regularity in $r=0$ requires $\re(p)\geq-|m|$. 
Then, the integrability and the absence of singularity at any finite $r$ and $z$ are only guaranteed  in the following four inequivalent cases:
\begin{align}
\notag
{\rm I)\,\,} &\im(q_0)>0\,,\,\, \im(q_1)=+\infty\,,\,\, \re(p)>-|m|-1
\\
\notag
{\rm II)\,\,} &\im(q_0)>0\,,\,\, \im(q_1)<0\,,\,\, p=2\ell \text{ with }\ell\in\mathbb N
\\
\notag
{\rm III)\,\,} &\im(q_0)>0\,,\,\, \im(q_1)=0\,,\,\, \re(p)\geq-|m|
\\
{\rm IV)\,\,} &\im(q_0)>0\,,\,\, \im(q_1)>0\,,\,\, \forall p\in \mathbb C
\label{conditions}
\end{align}
Other equivalent cases may be obtained by the symmetry
$(p,m, q_0,q_1)\rightarrow (-p-2|m|-2,m, q_1, q_0)$.

To further investigate the properties of the CiBs,
we now evaluate their free-space divergence, an important parameter for experimental
implementation and propagation through long distances.
For general light beam, the root mean square (rms) divergence $\thrms$  can be defined  as~\cite{mart93opl,stel00opc,padg14qph}
\beq
\label{thrms_def}
\thrms\equiv\lim_{z\rightarrow+\infty}\frac{\sigma_{\rm rms}(z)}{z}\,,
\eeq
where $\sigma^2_{\rm rms}(z)$  is  the second moment of the intensity $I(\xx)$:
\beq
\label{rms}
\sigma^2_{\rm rms}(z)=\int^{2\pi}_0\!\!\!\!\!\!\dd \phi \int^{+\infty}_0\!\!\!\!\!\!\!\!\! \dd r \,r\left[ r^2 \,I(\xx)\right]\,.
\eeq
By exploting  the expansion of the CiB in terms of the LG modes \eqref{LGexpansion} and  the relation 
$2\int\!{\rm d}\phi \int\!\dd r\,  r ^3 \,\text{LG}_{\ell,m}(\xx)  \text{LG}^*_{n,m}(\xx)=
{W^2(z)} [
B_{m,\ell} \delta _{n,\ell}
- C_{m,\ell} \delta _{n+1,\ell} 
-C^*_{m,n} \delta _{n,\ell+1} 
]$
with $B_{n,m}=|m|+2n+1$ and $C_{m,n}(z)=\sqrt{n (|m|+n)} \exp[2 i \zeta(z)]$ it is possible to give an explicit expression for $\sigma^2_{\rm rms}(z)$, namely:
\beq
\begin{aligned}
\frac{2\,\sigma^2_{\rm rms}(z+d_0)}{W^2(z)}=
1+|m|+\Phi_{p,m}^{(\xi)}+\re[X(z)]\,,
\end{aligned}
\eeq
where $X(z)=\xi (p -\Phi_{p,m}^{(\xi)})\exp[2i\zeta(z)]$ and
\beq
\Phi_{p,m}^{(\xi)}=
\frac{|p\, \xi|^2}{2+2|m|}\frac{_2F_1[1-\frac p2, 1-\frac{p^*}{2}, 2 + |m|,|\xi|^2]}{_2F_1[-\frac p2, -\frac{p^*}{2}, 1 + |m|,|\xi|^2]}\,.
\eeq
By plugging the above result into \eqref{thrms_def}, 
 the rms divergence of the CiBs may be expressed as
\beq
\label{thrms}
\begin{split}
\thrms=&\theta_0
\sqrt{1+|m|+\Phi_{p,m}^{(\xi)}-
\re\left[\xi  (p-\Phi_{p,m}^{(\xi)})\right]}\,.
\end{split}
\eeq
In the above equation $\theta_0={W_0}/{(\sqrt2 z_0)}$ is
the divergence of a Gaussian Beam (i.e. a CiB with $p=m=0$) with confocal parameter $z_0$
and beam waist $W_0=\sqrt{\lambda z_0/\pi}$.
As we will show in the following (see eq. \eqref{th_simple}), the divergence $\thrms$ is not finite when $\im(q_1)=0$ and $\re(p)=-|m|$.
Indeed, for such beams, at fixed $z$ and large $r$, the intensity fall-off as $r^{-4}$. While the field is thus square integrable,
the integral \eqref{rms} is divergent and the 
$\sigma^2_{\rm rms}(z)$ cannot be defined. 

\begin{figure}[t]
\includegraphics[width=8.5cm]{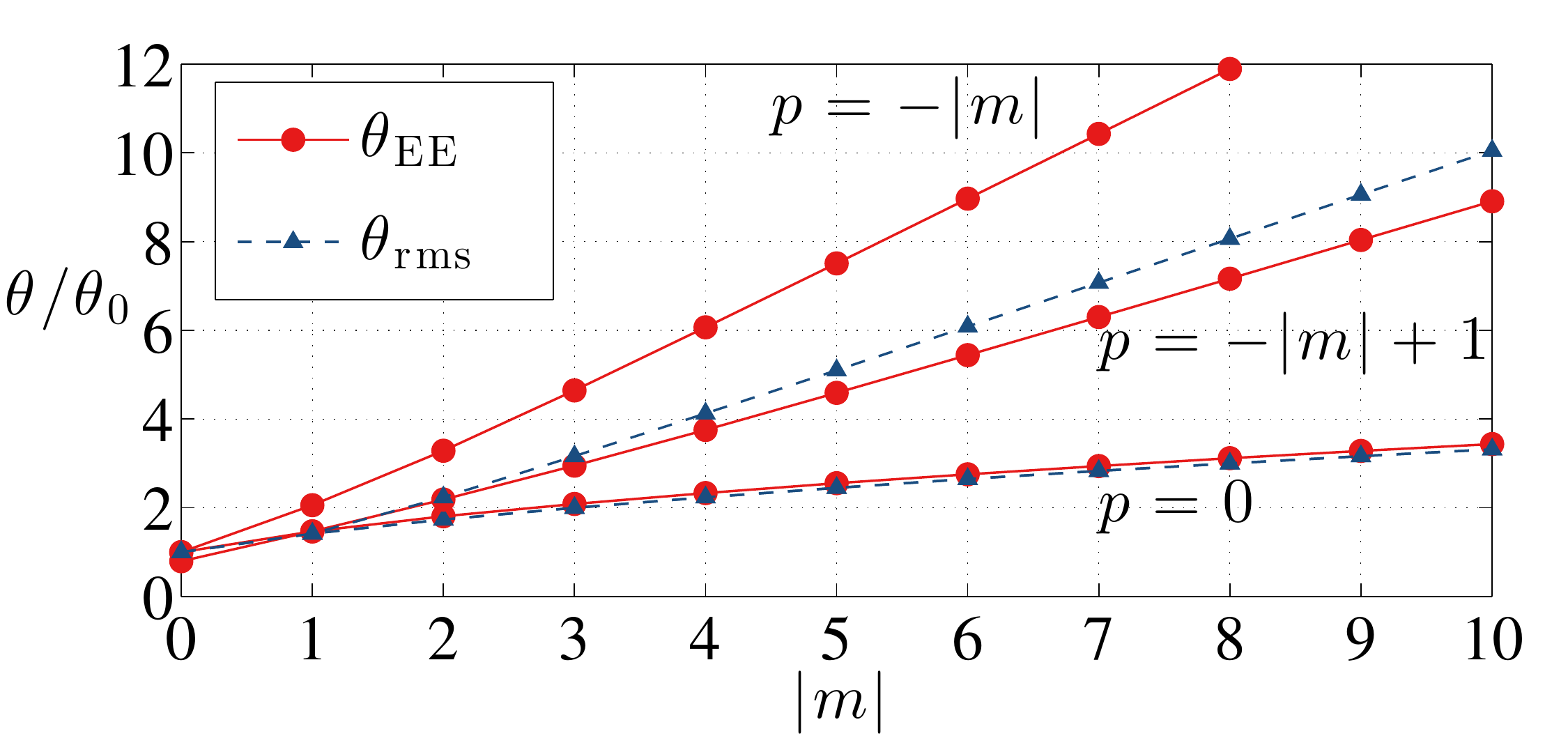}
\caption{Divergences $\thrms$ and $\thEE$
in function of $|m|$ for the CiB with $q_1=\re(q_0)$ and $p=0$, $p=-|m|+1$ or $p=-|m|$. When $p=-|m|$, only $\thEE$ can be defined.
The beam is only defined  for integer values of $m$ and lines represent just a guide for the overall trends.
Note that when $p=0$ the CiB$_{p,m}$ reduces to the Laguerre-Gauss mode LG$_{0,m}$.}
\label{divm}
\end{figure}
To avoid such problem, we may use an alternative definition of the divergence 
in terms of the so-called encircled-energy. It requires the determination of the radius $R_{\rm EE}(z)$  whose corresponding
circle centered on the beam axis contains a given amount of beam energy. Then, $R_{\rm EE}(z)$ must be calculated by 
the implicit relation $\int\dd\phi \int^{R_{\rm EE}(z)}_0\dd r\, r I(\xx)=I_0$, with 
$I_0$ a fixed constant. 
Given $R_{\rm EE}(z)$, the corresponding divergence $\thEE$ can be 
determined similarly to eq. \eqref{thrms}:
\beq
\theta_{\rm EE}=\lim_{z\rightarrow+\infty}\frac{R_{\rm EE}(z)}{z}\,.
\eeq
If we choose $I_0=1-1/e$ (chosen to achieve $\thEE=\thrms=\theta_0$ for the Gaussian beam),
the divergence $\thEE$ of a CiB can be obtained by the following implicit relation:
\beq
\label{thEE}
\int^{(\thEE/\theta_0)^2}_0\!\!\!\!\!\!e^{-t} t^{|m|}{\left|H_{p,m}(\frac{\xi t}{\xi+1})\right|}^2dt =I^{(\xi)}_{p,m}\,,
\eeq
with
\beq
I^{(\xi)}_{p,m}=\frac{e-1}{e}\frac{|m|!\, \Ftwoone}{|{(1+\xi)}^p|}\,,
\eeq
\noindent and $H_{p,m}(z)$ a shorthand for the Hypergeometric function $_1F_1(-p/2,|m|+1,z)$.
Relation \eqref{thEE} is obtained by inserting the expression of the CiB \eqref{cib} into the integral defining $R_{\rm EE}(z)$,
by changing variable into $r=W(z)\sqrt {t/2}$, by
taking the limit for $z\rightarrow+\infty$, and
by using the property 
$\lim_{z\rightarrow+\infty}{\sqrt 2 R_{\rm EE}(z)}/{W(z)}={\theta_{{\rm EE}}}/{\theta_0}$.

\begin{figure}
\includegraphics[width=8.5cm]{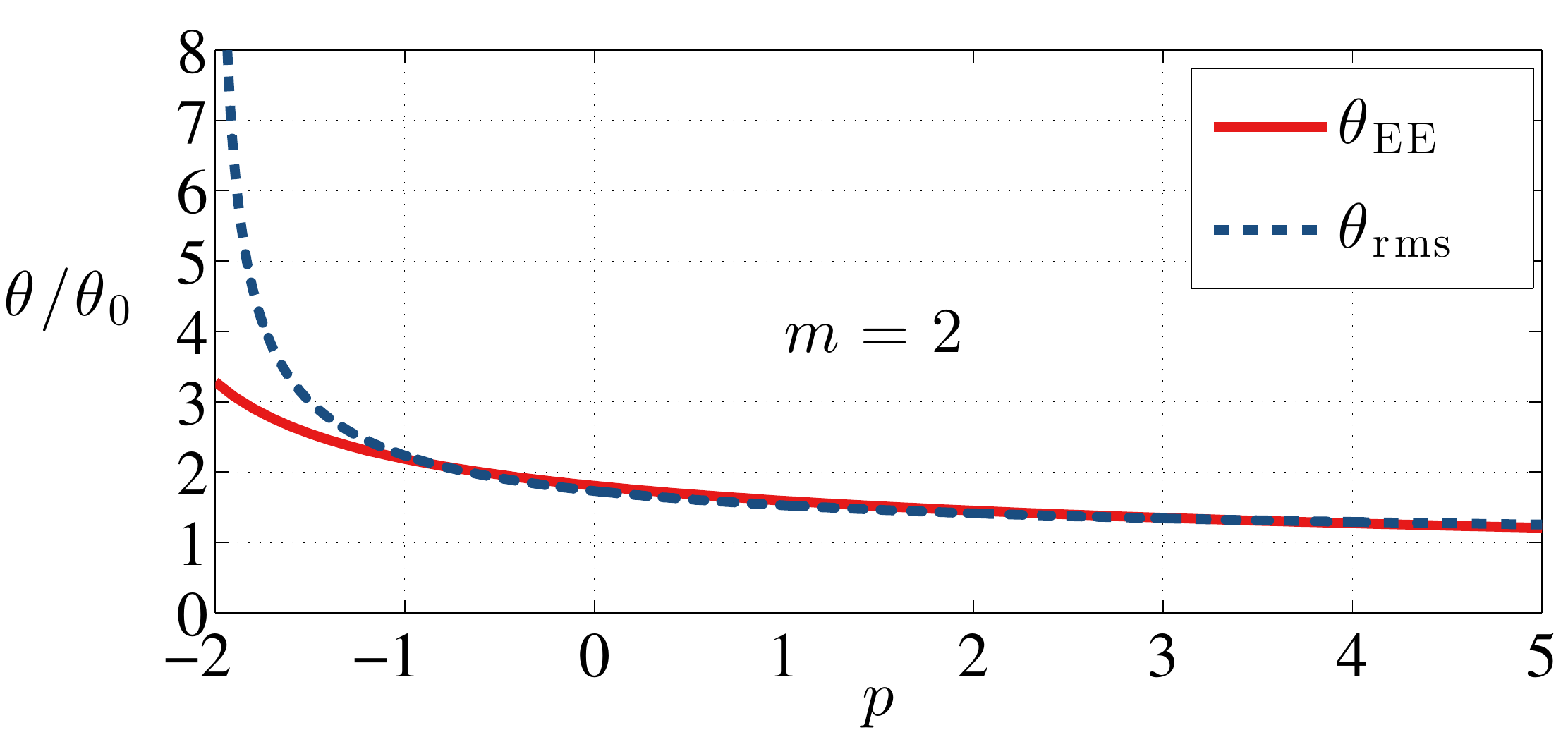}
\caption{Divergences $\thrms$ and $\thEE$
in function of $p$ for the CiB with $\im(p)=0$, $q_1=\re(q_0)$ and $m=2$. When $p$ is approaching to $-|m|$, the value $\thrms$ tends to infinity.}
\label{divp}
\end{figure}

To better understand the properties the CiBs, we now analyze their features for some particular values of the beam parameters and study in more
detail their divergences and their experimental generation.
Indeed, we would like to examine the CiBs with $|\xi|=1$ corresponding to $\im(q_1)=0$ or $\im(q_1)\rightarrow\infty$.
In this case, a simpler expression for the CiB is obtained.
Indeed, for $|\xi|=1$ we have 
$2\,\Phi_{p,m}^{(\xi)}={|p|^2}/{[\re(p)+|m|]}$
and the rms divergence and  normalization factor simplify to 
\begin{align}
\label{th_simple}
\notag
\thrms&=\theta_0
\sqrt{1+|m|-\re(p\, \xi)+\frac{|p|^2}{2}\frac{\re(\xi)+1}{\re(p)+|m|}}\,,
\\
\Ftwoone&=|m|!\,\,
\frac{\Gamma(|m|+1+\re(p))}{|\Gamma(|m|+1+\frac p2)|^2}\,.
\end{align}
As it is  evident from \eqref{th_simple}, when $\im(q_1)=0$ and $\re(p)=-|m|$ the rms divergence cannot be defined,
as already anticipated.
For such beams, the divergence can be only evaluated by $\thEE$. 
To better illustrate the behavior of the divergences, we show 
in Figure \ref{divm} the values of $\thrms$ and $\thEE$ in function of $m$
 for the CiB with $q_1=\re(q_0)$, such that $\xi=1$ and $\thrms=\theta_0\sqrt{1+{m^2}/{(p+|m|)}}$.
We show three different CiBs classes, corresponding to the $p$ parameter given by $p=0$, $p=-|m|+1$ or $p=-|m|$.
In the latter case ($p=-|m|$), as already discussed, only $\thEE$ can be shown since $\thrms$ in not defined.
The case $p=0$ corresponds to the LG mode LG$_{0,m}$.
Similarly, in {Figure} \ref{divp} we show the two divergences $\thrms$ and $\thEE$
in function of $p$ for the CiB with $\im(p)=0$,  $q_1=\re(q_0)$ and $m=2$. It is worth noticing that increasing the value of $p$ will reduce 
both divergences such that $\lim_{p\rightarrow+\infty}\theta=\theta_0$.

{The role of the imaginary part of the $p$ parameter can be appreciated when $q_1=-d_1$. In such case, 
at fixed $\re(p)$, the rms divergence is minimized for $\im(p)=(\re(p)+|m|)\frac{d_1-d_0}{z_0}$ and
 becomes a decreasing function of $\re(p)$, namely 
 $\thrms=\theta_0
\sqrt{1+\frac{m^2}{\re(p)+|m|}\frac{z^2_0}{z^2_0+(d_0-d_1)^2}}$. Then, when $-q_1=d_1\neq d_0$, 
a complex value of $p$ is required  to minimize the
beam divergence.}

We finally analyze a possible experimental generation of such subclass of CiBs.
The cases $q_1=\re(q_0)=-d_0$ and $\im(q_1)\rightarrow\infty$ respectively correspond to the HyGG and HyGG-II modes introduced in
\cite{kari07opl,kari09opl}. The general case $q_1=-d_1$, that we call generalized Hypergeometric-Gaussian beam (gHyGG), 
is particularly interesting from the point of view of experimental generation. 
Indeed, the behavior of the field for $z\rightarrow d_1$ gives a clear hint for its possible experimental generation. For $q_1=-d_1$ we have
\beq
\label{pupil_limit}
\lim_{z\rightarrow d_1}\CiB
\propto
e^{-\frac{ikr^2}{2(d_1-d_0+iz_0)}}\frac{r^{p+|m|}e^{im\phi}}{d_1-d_0+iz_0}\,.
\eeq
This feature is very useful because the r.h.s. of eq. \eqref{pupil_limit} can be generated by applying the singular phase factor 
$\exp(im\phi)$
and a polynomial transmittance profile of the order $p+|m|$
to Gaussian  (TEM$_{00}$) beam at a distance $d_1-d_0$ from its waist. In particular, the $p=-|m|$ modes are simply
generated by applying the phase factor $\exp(im\phi)$ to a Gaussian beam at a distance $d_1-d_0$ from its waist plane.
Such phase factor can be experimentally implemented by q-plates~\cite{marr06prl} or phase-plates~\cite{beij94opc}.

In the present letter we studied the properties of the Circular-Beams, a general solution
of the paraxial wave equation with OAM depending on three complex parameters
$q_0$, $q_1$ and $p$ and an integer parameter $m$, related to the content of OAM. The allowed values of the beam
parameters are given in eq. \eqref{conditions} and the corresponding ones obtained by the symmetry
$(p,m, q_0,q_1)\rightarrow (-p-2|m|-2,m, q_1, q_0)$.
We  derived the normalization of the CiBs and their expansion  in terms
of Laguerre-Gauss modes, a standard OAM basis (see eq. \eqref{LGexpansion}). 
Such expansion allowed us to study the
divergence of the CiBs in free-space propagation. We defined the divergence by two methods, the rms (see eq. \eqref{thrms}) and
the encircled-energy (see eq. \eqref{thEE}) and studied their behavior for particular values of the beam parameters.
We finally suggested the experimental generation of a subclass of  CiBs by looking at their behavior in
a particular transverse plane, see \eqref{pupil_limit}.
{Our achievements, providing the estimation of the beam divergence for the CiBs (the most general beam with OAM known so far),
may have potential application for OAM transportation in free space, such as multi-channel/multiplexing classical
or quantum communication~\cite{tamb12njp, bozi13sci,vall14prl}.}

We thanks Paolo Villoresi of the University of Padova for suggestions and useful discussions.
Our work was supported by the 
``Progetto di Ateneo PRAT 2013, {\it OAM in free space: a new resource for QKD} (CPDA138592)"
of the University of Padova.

%
%
%
%
%
%

\end{document}